# On the Conformal Forms of the Robertson-Walker Metric

Running title: Conformal Forms of the RW Metric

M Ibison [a)]

*Institute for Advanced Studies at Austin,*

*11855 Research Boulevard, Austin TX 78759-2443, USA*

All possible transformations from the Robertson-Walker metric to those conformal to the Lorentz-Minkowski form are derived. It is demonstrated that the commonly known family of transformations and associated conformal factors are not exhaustive and that there exists another relatively less well known family of transformations with a different conformal factor in the particular case that $\kappa = -1$. Simplified conformal factors are derived for the special case of maximally-symmetric spacetimes. The full set of all possible cosmologically-compatible conformal forms is presented as a comprehensive table. A product of the analysis is the determination of the set-theoretical relationships between the maximally symmetric spacetimes, the Robertson-Walker spacetimes, and functionally more general spacetimes. The analysis is preceded by a short historical review of the application of conformal metrics to Cosmology.

## I. INTRODUCTION

The fact that the Weyl Tensor for all Robertson-Walker Cosmological metrics vanishes meets the necessary and sufficient condition that a conformal form of these metrics must also exist [1]. A useful property of a conformal metric is that it leaves Maxwell's equations unchanged

[a)] Electronic mail: ibison@earthtech.org

from their form in Minkowski spacetime [2,3], making EM calculations in conformally-represented spacetimes particularly simple. Despite this, the conformal metrics corresponding to the curved space - $K = \pm 1$ - RW [b)] spacetimes are not given in major texts on Cosmology and General Relativity. The books by Tolman [4], Weinberg [1], Hawking and Ellis [5], Misner, Thorne and Wheeler [6], Birrell and Davies [2], Wald [3], Schutz [7], Peebles [8], Harrison [9], Roos [10], Dodelson [11], Carroll [12], Hobson et al [13], and Bergström and Goobar [14] do not mention conformal forms of the $K = \pm 1$ RW metrics. Where they discuss conformal maps of these spacetimes, they do so with metrics that are not conformal to the Lorentz-Minkowski metric - which they are able to do because it is sufficient (but not necessary) for the purposes of constructing conformal diagrams that the metric be expressible in the form [12]

$$\mathrm{d}s^2 = f^2(t,r)\left(\mathrm{d}t^2 - \mathrm{d}r^2\right) + g^2(r)\,\mathrm{d}\Omega^2 . \tag{1}$$

Landau and Lifshitz [15] give, without derivation, the 'exponential' transformation (see Eq. (58) below) from RW $K = -1$ metric to conformal form, but do not report that in this case the conformal factor is then a single parameter function of the Minkowski square. And they do not give the better-known 'hyperbolic' transformation valid for all three $K = \pm 1$ RW metrics. Peacock [16] mentions that the $K = \pm 1$ RW metrics can be put into conformal form though does not give any details. Lightman et al [17] sketch a derivation of the hyperbolic transformation, though they do not actually give the conformal factor. Their analysis is not exhaustive however; it fails to find the exponential transformation and corresponding conformal factor valid for the $K = -1$ RW metric. (We reproduce their method in Section A5, showing where the additional solution goes missing.) Though Stephani et al [18] give none of the conformal forms of the RW metrics, they acknowledge the existence of such forms in connection with Tauber's publication [19]. Stephani [20] states that all RW metrics can be put into conformal form, and explicitly gives the hyperbolic transformation for the case $K = +1$ RW metric, but does not mention the exponential form.

In the journal literature it appears that Infeld and Schild [21], using a kinematic analysis, were the first to report that there are two distinct families of conformal metrics (namely the hyperbolic and the exponential) that can be mapped onto the same RW metric when $K = -1$. (Strangely, they appear to confound physically different metrics - not related by a coordinate transformation -

---

[b)] With apologies to Friedmann and Lemaitre, RW is used throughout as shorthand for FLRW.

with physically different universes – i.e. wherein red-shifts do or do not occur.) The exponential transformation had been obtained earlier by Walker [22]. In a subsequent publication Infeld and Schild [23] showed how to transform the Maxwell and Dirac equations from Lorentz-Minkowski form to curved spacetime expressed in conformal coordinates. Much later Tauber [19] used the results of Infeld and Schild [21] to derive the associated Friedmann equation for the conformal factor, which he solved for some special cases of pressure and energy density. The conformal forms of the RW metrics were the focus of several works by Endean [24-27]. These built on Tauber's analysis, but included a claim that the observation data had been consistently incorrectly interpreted through the lens of the RW metrics, and that for example the cosmological age-problem (of that time) would be solved if properly treated with the conformal metrics. But in the light of the general coordinate transformation invariance of GR, this claim could be true only if the consensus application of GR to the observational data at that time were in error. In a subsequent analysis of Endean's works Querella [28] showed his conclusions rested on a non-standard interpretation of coordinate times and distances in the conformal coordinate system. Herrero and Morales [29] applied the kinematic approach of Infeld and Schild to decide constraints on a suitable conformal Killing vector field which they used to determine the cosmologically-compatible conformal factors. However, their approach is sufficiently different to that adopted here that it is not immediately clear if their result encompass all the cases given here, including in particular the exponential transformation from the RW $K = -1$ metric. Sopuerta [30], and Keane and Barrett [31] derive transformations from the RW metrics to conformal form without special treatment of the $K = -1$ case.

During the revision of this manuscript Iihoshi et al [32] posted a paper giving the full solution, i.e. compatible with (A23), without proof, to the functional equation (17). They observe that by suitable choice of integration constants - which can be shown are equivalent to those in (A24) - the transformation becomes of the exponential type, only in the case $K = -1$. They do not report on the fact that for this solution the conformal factor reduces to a single parameter function of the Minkowski square (Eq. (58) below).

The hyperbolic transformation is the focus of Sec. IV and is valid for all RW spacetimes. In Secs. V and VI the results of Sec. IV are used to give explicit conformal forms of the maximally symmetric spacetimes, showing, at the same time, some set-theoretical relationships between them. The conformal form of the specifically flat-space RW spacetimes is derived in Sec. VII, based upon a limiting form of the RW-compliant conformal factor of Sec. IV. In Sec. VIII is presented the second family of transformations characterized by an exponential transformation of the RW coordinates, which is specific to the case $K = -1$. All these forms of the RW metric are collected together in Table I. The set-theoretical relationships showing how the RW metrics overlap in their capacity to describe an underlying spacetime are given in Fig. I.

In this document ‘spacetime’ is shorthand for pseudo-Riemannian manifold. The line element $\mathrm{d}s^2 = \mathrm{d}x^2$ and associated metric will be referred to as ‘Lorentz-Minkowski’, and the underlying spacetime ‘Minkowski’. No distinction is made between different topologies that are otherwise locally equivalent through a coordinate transformation. All metrics under consideration here are diagonal, and are most conveniently expressed as a line element. Discussion of the extensibility of a coordinate system and its covering of a given spacetime is mostly avoided.

## II. Traditional forms of the Robertson-Walker metric

Robertson-Walker spacetimes [33,34] were first deduced from the ‘Cosmological Principle’ without explicit reference to a stress energy tensor or to GR. The principle states that whilst permitting the possibility of variation in time, in the large the universe should otherwise look the same to (an appropriately defined class of) observers everywhere. From this was deduced the invariant interval

$$\mathrm{d}s^2 = \mathrm{d}t_c^2 - a^2\left(t_c\right)\left(\mathrm{d}r^2 + s^2\left(r\right)\mathrm{d}\Omega^2\right) \tag{2}$$

where

$$s^2\left(r\right) = \begin{cases} \left(\sin\left(kr\right)/k\right)^2 & \textit{spherical} \\ r^2 & \textit{flat} \\ \left(\sinh\left(kr\right)/k\right)^2 & \textit{hyperbolic} \end{cases} \tag{3}$$

and where the scale factor $a\left(t_c\right)$ is an arbitrary function of the cosmological time - which is also the proper time of the fundamental observer. The three cases - spherical, flat, and hyperbolic - refer to the curvature of space, but, as will be seen, do not single out any particular spacetime and to some degree overlap. These three can be combined notationally into

$$\mathrm{d}s^2 = \mathrm{d}t_c^2 - a^2\left(t_c\right)\left(\mathrm{d}r^2 + \frac{R_0^2}{K}\sin^2\left(\sqrt{K}r / R_0\right)\mathrm{d}\Omega^2\right) \tag{4}$$

where $K \in [1,0,-1]$ covers the three cases (3), and $R_0$ is some fixed distance. It will be convenient to write subsequent expressions for the line element in normalized coordinates; the possibility of a further real linear transformation of the coordinates therein, $x \to x' = ax + b$ with $a$ and $b$ real, will then be understood. In such terms (4) becomes

$$\mathrm{d}s^2 = \mathrm{d}t_c^2 - a^2(t_c)\left(\mathrm{d}r^2 + S_K^2(r)\,\mathrm{d}\Omega^2\right) \tag{5}$$

where the definition

$$S_K(r) = \sin\left(\sqrt{K}r\right)/\sqrt{K}; \quad K = [1,0,-1]$$

will also be useful elsewhere. In (5) the scale factor $a$, the cosmological time $t_c$, and the increment d$s$ are to be regarded as similarly normalized with respect to a fixed distance [c)].

The coordinate change $\bar{r} = 2\tan\left(\sqrt{K}r/2\right)/\sqrt{K}$ takes (5) into the isotropic form [16]

$$\mathrm{d}s^2 = \mathrm{d}t_c{}^2 - \frac{a^2(\tau)}{\left(1 + K\bar{r}^2/4\right)^2}\left(\mathrm{d}\bar{r}^2 + \bar{r}^2\mathrm{d}\Omega^2\right). \tag{6}$$

(In this case the factor of ¼ multiplying $\bar{r}^2$ can be removed by rescaling $a^2(\tau)$ and the coordinates. If $a(t_c)$ is not arbitrary however – as in the case of a maximally symmetric spacetime – then removing the factor ¼ from the denominator by implicit redefinition of the coordinates will, in general, have consequences elsewhere in the expression for the line element.) A more common form used in magnitude distance calculations is related to (6) by the coordinate change $\tilde{r} = S_K(r)$, which gives

$$\mathrm{d}s^2 = \mathrm{d}t_c^2 - a^2(t_c)\left(\frac{\mathrm{d}\tilde{r}^2}{1 - K\tilde{r}^2} + \tilde{r}^2\mathrm{d}\Omega^2\right). \tag{7}$$

Note that the new radii are real for all $K \in [1,0,-1]$. All three forms - (5), (6), (7) - can be recast with a conformal time $t$ defined by [d)] $\mathrm{d}t_c = a(t_c)\mathrm{d}t$ so that for example (5) becomes [e)]

---

[c)] An alternative is to regard the scale factor $a(t_c)$ and the cosmological time $t_c$ in this particular line element as having units of length ($c$ = 1 throughout), and have only the radial coordinate normalized. In that case the increment d$s$ also has units of length.

$$\mathrm{d}s^2 = a^2(t)\left(\mathrm{d}t^2 - \mathrm{d}r^2 - S_K^2(r)\mathrm{d}\Omega^2\right). \tag{8}$$

The collection of Robertson-Walker spacetimes described above will be denoted by $\{\mathrm{RW}_K(t)\}$, which includes all three cases $K = [-1,0,1]$ and where $(t)$ stands for the functional degree of freedom that in this case is the RW scale factor $a(t)$. For example the hyperbolic space form of the RW metric generates spacetimes $\{\mathrm{RW}_{-1}(t)\}$. Allowing for some overlap between the subsets, one has

$$\{\mathrm{RW}_K(t)\} = \{\mathrm{RW}_{-1}(t)\} \cup \{\mathrm{RW}_0(t)\} \cup \{\mathrm{RW}_{+1}(t)\}.$$

In the following it will also be useful to refer also to the spacetimes that can be written in any of the equivalent terms

$$\begin{aligned}
\mathrm{d}s^2 &= \mathrm{d}t^2 - a^2(t)s^2(r)\left(\mathrm{d}r^2 + r^2\mathrm{d}\Omega^2\right) \\
&= a^2(t)\left(\mathrm{d}t^2 - s^2(r)\left(\mathrm{d}r^2 + r^2\mathrm{d}\Omega^2\right)\right) \\
&= \mathrm{d}t^2 - a^2(t)\left(s^2(r)\mathrm{d}r^2 + r^2\mathrm{d}\Omega^2\right) \\
&= a^2(t)\left(\mathrm{d}t^2 - s^2(r)\mathrm{d}r^2 - r^2\mathrm{d}\Omega^2\right) \\
&= \mathrm{d}t^2 - a^2(t)\left(\mathrm{d}r^2 + s^2(r)\mathrm{d}\Omega^2\right) \\
&= a^2(t)\left(\mathrm{d}t^2 - \mathrm{d}r^2 - s^2(r)\mathrm{d}\Omega^2\right)
\end{aligned} \qquad \text{(9 a,b,c,d,e,f)}$$

where both $s(r)$ and $a(t)$ are arbitrary functions (see for example [16]). In these spacetimes the variation in time is decoupled from the spatial variation. The first and second forms are isotropic. In the second and fourth forms the scale factor applies to all components of the interval. Henceforth these spacetimes will be denoted OTI standing for ‘orthogonal time isotropic’ and more formally by $\{\mathrm{OTI}(t)(r)\}$ where $(t)$ and $(r)$ stand respectively for the functional degrees of freedom and $s^2(r)$ in any of the equivalent forms in (9).

---

[d)] We choose not to use the traditional symbol for the conformal time $\eta$ because later we will want to refer collectively to all four coordinates in which the metric is conformal by just ‘*x*’, in which, traditionally, the $0^{\text{th}}$ component is synonymous with *t*.

[e)] A function will sometimes be implicitly redefined after making a coordinate change; the scale factor *a* in (8) is not the same function of its argument as the scale factor in (5).

## III. Robertson-Walker metric in Lorentz-conformal coordinates

In this section is sought the coordinate transformation that takes the OTI line element (9f) into the form

$$\mathrm{d}s^2 = A^2\left(T,R\right)\left(\mathrm{d}T^2 - \mathrm{d}R^2 - R^2\mathrm{d}\Omega^2\right) = A^2\left(T,R\right)\mathrm{d}X^2 \tag{10}$$

where $\mathrm{d}X^2$ is the invariant interval of Minkowski space-time in Lorentz-Minkowski coordinates:

$$\mathrm{d}X^2 \equiv \mathrm{d}T^2 - \mathrm{d}\mathbf{X}\bullet\mathrm{d}\mathbf{X}.$$

Where it is convenient, all spacetimes that can be written in the (conformal) form (10) will be referred to by $\left\{\text{conformal}\left(t,r\right)\right\}$, the parameters $t$ and $r$ being necessary to distinguish these from more restricted spacetimes wherein the conformal factor is a function of just one generalized coordinate. The relationship between the new and old coordinates is

$$\mathrm{d}T = T_t\mathrm{d}t + T_r\mathrm{d}r, \quad \mathrm{d}R = R_t\mathrm{d}t + R_r\mathrm{d}r.$$

Inserting this into (10) and equating with (9f) gives

$$T_t^2 - R_t^2 = R_r^2 - T_r^2 = R^2/s^2 = a^2/A^2 \tag{11}$$

and

$$T_tT_r - R_tR_r = 0. \tag{12}$$

The assumption is that $a$ and $A$ are positive. The first two of Eqs. (11) give

$$T_t^2 + T_r^2 = R_t^2 + R_r^2$$

which with (12) gives

$$\left(T_t + T_r\right)^2 = \left(R_t + R_r\right)^2, \quad \left(T_t - T_r\right)^2 = \left(R_t - R_r\right)^2$$
$$\Rightarrow \quad T_t + T_r = \varepsilon_1\left(R_t + R_r\right), \quad T_t - T_r = \varepsilon_2\left(R_t - R_r\right)$$

where the $\varepsilon_i$ are $\pm 1$. If $\varepsilon_1 = \varepsilon_2$ then $T_t = \varepsilon_1 R_t$ and (11) would then require $a = 0$. Therefore the $\varepsilon_i$ must be different, giving

$$T_t = \varepsilon R_r, \quad T_r = \varepsilon R_t \tag{13}$$

where $\varepsilon = \pm 1$. These give

$$T_{rr} = T_{tt} \Rightarrow T = f\left(t+r\right) + g\left(t-r\right) \tag{14}$$

where $f$ and $g$ are arbitrary functions. Putting this into (13) gives

$$\varepsilon R = f\left(t+r\right) - g\left(t-r\right) \tag{15}$$

where an arbitrary constant of integration has been absorbed into the definitions of *f* and *g*. Inserting these into Eq. (11) gives the functional differential equation

$$\left(f\left(t+r\right) - g\left(t-r\right)\right)^2 = 4s^2\left(r\right)\dot{f}\left(t+r\right)\dot{g}\left(t-r\right). \tag{16}$$

It is helpful to go to sum and difference (i.e. light-cone based) coordinates $t+r=u,\quad t-r=v$ whereupon (16) becomes

$$\frac{4\dot{f}\left(u\right)\dot{g}\left(v\right)}{\left(f\left(u\right)-g\left(v\right)\right)^2} = \frac{1}{s^2\left(\left(u-v\right)/2\right)}. \tag{17}$$

Noticing that the left hand side can be written as a perfect differential

$$\frac{4\dot{f}\left(u\right)\dot{g}\left(v\right)}{\left(f\left(u\right)-g\left(v\right)\right)^2} = -4\frac{\partial}{\partial u}\frac{\dot{g}\left(v\right)}{f\left(u\right)-g\left(v\right)} = 4\frac{\partial^2}{\partial u\partial v}\log\left(f\left(u\right)-g\left(v\right)\right),$$

It follows that the general solution of (17) is

$$\log\left(f\left(u\right)-g\left(v\right)\right) = -\int^{\frac{u-v}{2}} \mathrm{d}r'\int^{r'}\mathrm{d}r'' s^{-2}\left(r''\right) + p\left(u\right) + q\left(v\right), \tag{18}$$

where $f\left(u\right), g\left(v\right), p\left(u\right), q\left(v\right)$ are otherwise arbitrary functions. Eq. (18) can be written

$$f\left(u\right) - g\left(v\right) = \phi\left(u\right)\gamma\left(v\right)\sigma\left(u-v\right) \tag{19}$$

where $f, g, \phi = e^p, \gamma = e^q$ are otherwise arbitrary functions, and $\sigma$ is given by

$$\sigma\left(w\right) = \exp\left(-\int^{w/2}\mathrm{d}w'\int^{w'}\mathrm{d}w'' s^{-2}\left(w''\right)\right). \tag{20}$$

Lightman et al [17] derive the functional equation (17) (Eq. 6 in Chapter 19 Solutions) with an *a priori* assumption that $f\left(x\right) = g\left(x\right)$. They sketch a solution to this equation that bypasses the introduction of the intermediate functions $\phi\left(u\right), \gamma\left(v\right)$ as they appear in (19). However it turns out there are two distinct solutions, namely (A6) and (A20), the first of which is overlooked by the Lightman method. A full derivation of both solutions is given in the appendix, along with the connection with the Lightman method. The next section deals with the transformation based on (A20). Sections V, VI, & VII consider some special cases. Section VIII deals with the transformation based on (A6).

## IV. Hyperbolic transformation to conformal form

From Eqs. (A20), (14) and (15) it is concluded that the associated coordinate transformation that takes (9f) into the form (10) is

$$f\left(u\right)=B+L\tanh\left(\kappa u-\chi_1\right),\quad g\left(v\right)=C+L\tanh\left(\kappa v-\chi_2\right) \tag{21}$$

and therefore

$$\begin{aligned} T &= B+C+L\Big(\tanh\big(\kappa\left(t+r\right)-\chi_1\big)+\tanh\big(\kappa\left(t-r\right)-\chi_2\big)\Big) \\ \varepsilon R &= B-C+L\Big(\tanh\big(\kappa\left(t+r\right)-\chi_1\big)-\tanh\big(\kappa\left(t-r\right)-\chi_2\big)\Big) \end{aligned} \tag{22}$$

where $B,C,L,\kappa,\chi_i$ are constants. $L$ sets the length scale for the ($T$,$R$) coordinate system, and $1/\kappa$ sets the length scale for the ($t$,$r$) coordinate system. The sum and differences of the $\chi_i$ are constant offsets to the origin of the coordinates $t$ and $r$, and $B$ and $C$ are constant offsets to the origin of the coordinates $T$ and $R$. For the sake of notational simplicity, Eqs. (21) and (23) will be re-expressed in dimensionless coordinates with no explicit offset and with the implicit assumption that in any result the coordinates can be linearly rescaled:

$$\begin{gathered} f\left(u\right)=T_K\left(u/2\right),\quad g\left(v\right)=T_K\left(v/2\right) \\ T=T_K\left(\left(t+r\right)/2\right)+T_K\left(\left(t-r\right)/2\right),\quad \varepsilon R=T_K\left(\left(t+r\right)/2\right)-T_K\left(\left(t-r\right)/2\right) \end{gathered} \tag{24}$$

where the definition

$$T_K\left(x\right)\equiv\tan\left(\sqrt{K}x\right)/\sqrt{K};\quad K\in\left[1,0,-1\right]$$

will also be useful elsewhere. The denominator of $\sqrt{K}$ ensures that the coordinates are real for any $K$. Eq. (24) is the transformation derived by Lightman et al [17] and corresponds to $\kappa=\mathrm{i}\sqrt{K}/2,\quad L=-\mathrm{i}/\sqrt{K}$ in (21).

Eq. (16) with (24) gives

$$s^2\left(r\right)=S_K^2\left(r\right),$$

just as it appears in (8). Remarkably therefore, it turns out that the spatial curvature functions $S_K^2\left(r\right)$ stipulated by the RW metrics are the *only* functions that permit transformation via (A20) from the OTI form (9f) (i.e. with $s^2\left(r\right)$ arbitrary) to the conformal form. It turns out that this is true also of the exponential solution (A6) discussed in the next section. Anticipating that result, one has [f)]

$$\left\{\mathrm{RW}_K\left(t\right)\right\} = \left\{\mathrm{OTI}\left(t\right)\left(r\right)\right\} \cap \left\{\mathrm{conformal}\left(t,r\right)\right\} \tag{25}$$

where here $\left(t,r\right)$ denotes the full functional dependency of the conformal scale factor. The relation (25) is expressed in Fig. I.

From (11), (14) and (15) it is inferred that the conformal factor is

$$A^2\left(T,R\right) = \frac{a^2\left(t\right)}{4\dot{f}\left(t+r\right)\dot{g}\left(t-r\right)}. \tag{26}$$

Employing (24) this becomes

$$\begin{aligned} A^2\left(T,R\right) &= \frac{a^2\left(t\right)}{\left(1+\tan^2\left(\sqrt{K}\left(t+r\right)/2\right)\right)\left(1+\tan^2\left(\sqrt{K}\left(t-r\right)/2\right)\right)} \\ &= \frac{a^2\left(t\right)}{\left(1+\left(\sqrt{K}\left(T+\varepsilon R\right)/2\right)^2\right)\left(1+\left(\sqrt{K}\left(T-\varepsilon R\right)/2\right)^2\right)} \\ &= \frac{a^2\left(t\right)}{1+K^2X^4/16+\left(K/2\right)\left(T^2+R^2\right)} \\ &= \frac{a^2\left(t\right)}{\left(1-KX^2/4\right)^2+KT^2} = \frac{a^2\left(t\right)}{\left(1+KX^2/4\right)^2+KR^2} \end{aligned} \tag{27}$$

where $X^2 \equiv T^2 - R^2 = T^2 - \mathbf{X}\bullet\mathbf{X}$. The inverse relations between ($t$,$r$) and ($T$,$R$) and can be deduced from (24). Specifically

$$\begin{aligned} &\sqrt{K}\left(T \pm \varepsilon R\right)/2 = \tan\left(\sqrt{K}\left(t \pm r\right)/2\right) \\ &\Rightarrow t = \frac{\tan^{-1}\left(\sqrt{K}\left(T+\varepsilon R\right)/2\right)+\tan^{-1}\left(\sqrt{K}\left(T-\varepsilon R\right)/2\right)}{\sqrt{K}} = \frac{1}{\sqrt{K}}\tan^{-1}\frac{\sqrt{K}T}{1-KX^2/4}. \end{aligned}$$

Putting this into (27), the conformal scale factor becomes

---

[f)] Here and elsewhere the members of the set are the distinct spacetimes that are not equivalent under an allowed general coordinate transformation (one that is real and preserves the number of positive and negative eigenvalues of the metric). In these expressions involving different families of spacetimes possible differences in the coverage of a spacetime by a coordinate system (chart) are ignored. Relations such as (25) may therefore be valid only over a restricted 'patch'.

$$A^2(T,R) = \frac{1}{\left(1-KX^2/4\right)^2 + KT^2} a^2 \left( \frac{1}{\sqrt{K}} \tan^{-1} \frac{\sqrt{K}T}{1-KX^2/4} \right). \tag{28}$$

It may be observed that the scale factor can be factorized as the product of two one-parameter functions in two different ways:

$$A^2(T,R) = B^2(z_K;K)/T^2 = C^2(z_K;K)/\left(1-KX^2/4\right)^2 \tag{29}$$

where

$$B^2(z_K;K) \equiv \frac{a^2\left(\tan^{-1}\left(\sqrt{K}z_K\right)/\sqrt{K}\right)}{K+1/z_K^2}, \quad C^2(z_K;K) \equiv \frac{a^2\left(\tan^{-1}\left(\sqrt{K}z_K\right)/\sqrt{K}\right)}{1+Kz_K^2} \tag{30}$$

are arbitrary functions, and

$$z_K = \frac{T}{1-KX^2/4} \tag{31}$$

is real. (It can be shown that there is no similar factorization involving just *R* and *X*.)

In summary, using the transformation (24) all RW spacetimes can be written in conformal Lorentz-Minkowski form with scale factor given by (29). Eq. (28) establishes the relationship between the conformal scale factor and the scale factor of the traditional RW form (9f). Symbolically, this relationship can be written

$$\left\{\mathrm{RW}_K(t)\right\} = \left\{\mathrm{conformal} \frac{1}{t^2}\left(\frac{t}{1-Kx^2}\right)\right\} = \left\{\mathrm{conformal} \frac{1}{1-Kx^2}\left(\frac{t}{1-Kx^2}\right)\right\} \tag{32}$$

where the parentheses denote a functional degree of freedom in the metric (in this case the conformal scale factor). The leading factor is not a functional degree of freedom but a fixed coefficient. The factors of ¼ multiplying $x^2$ appearing in (29) and (31) can be removed by rescaling the coordinates, redefining the functions *B* and *C*, and absorbing an overall factor into the interval d*s*.

## V. Conformal and RW forms of de Sitter spacetime

A restricted case of the conformal line element is that the conformal factor is a function of time only. Clearly, these must be identical with the spacetimes $\left\{\mathrm{RW}_0(t)\right\}$. According to (29) for this restricted class of spacetimes one must be able to write

$$B\left(\frac{T}{1-KX^2/4};K\right) = f(T;K)$$

for any function *f*. Since this must be true for all *R* and therefore all *X* independent of *T*, it follows that either $K=0$ or $B=\text{constant}$. The first of these is the same as (8) with $K=0$, which is just

$$\mathrm{d}s^2 = a^2\left(t\right)\mathrm{d}x^2 . \tag{33}$$

In the second case, up to an arbitrary global scaling of the coordinates, the line element is just

$$\mathrm{d}s^2 = \mathrm{d}X^2/T^2 \tag{34}$$

which is the de Sitter spacetime in conformal coordinates. Note that this form can be achieved by any of $K=\left[-1,0,1\right]$, which of course includes the case (33) wherein $a^2\left(t\right)=1/t^2$ in the original coordinates. In summary:

$$\left\{\text{de Sitter}\right\}\subseteq\left\{\text{RW}_{-1}\left(t\right)\right\}\cap\left\{\text{RW}_{0}\left(t\right)\right\}\cap\left\{\text{RW}_{+1}\left(t\right)\right\}.$$

In fact, it is possible to show that the de Sitter spacetime is the *only* spacetime which can be expressed in all three Robertson Walker spacetimes, whereupon

$$\left\{\text{RW}_{-1}\left(t\right)\right\}\cap\left\{\text{RW}_{0}\left(t\right)\right\}\cap\left\{\text{RW}_{+1}\left(t\right)\right\}=\left\{\text{de Sitter}\right\} \tag{35}$$

i.e. the de Sitter spacetime can be uniquely defined this way. This relation is illustrated in Fig 1.

The (three) line elements (for each $K=\left[-1,0,1\right]$) having the form (8) that transform to (34) can be determined from setting *B* as defined in (30) to a constant (which can be set to 1):

$$\frac{a^2\left(\tan^{-1}\left(\sqrt{K}z_K\right)/\sqrt{K}\right)}{K+1/z_K^2}=1\Rightarrow a^2\left(t\right)=K\left(1+\cot^2\sqrt{K}t\right)=S_K^{-2}\left(t\right).$$

The de Sitter line element in these coordinates is therefore

$$\mathrm{d}s^2 = S_K^{-2}\left(t\right)\left(\mathrm{d}t^2-\mathrm{d}r^2-S_K^2\left(r\right)\mathrm{d}\Omega^2\right). \tag{36}$$

Explicitly:

$$\mathrm{d}s^2=\begin{cases}\left(\mathrm{d}t^2-\mathrm{d}r^2+\sinh^2 r\,\mathrm{d}\Omega^2\right)/\sinh^2 t & K=-1\\ \mathrm{d}x^2/t^2 & K=0\\ \left(\mathrm{d}t^2-\mathrm{d}r^2+\sin^2 r\,\mathrm{d}\Omega^2\right)/\sin^2 t & K=+1\end{cases}. \tag{37}$$

The second case is just (34) due to the fact that $K=0$ causes the transformation to become the trivial $T=t,\ R=r$. Eq. (36) can be cast into the RW form (2) with the transformation

$$t_c=-\int\mathrm{d}tS_K^{-1}\left(t\right)=-\int\mathrm{d}t\frac{\sqrt{K}}{\sin\sqrt{K}t}=-\log\frac{\tan\sqrt{K}t/2}{\sqrt{K}} \tag{38}$$

where the divisor ensures that $t$ is real when $K=-1$, and finite when $K=0$. Therefore

$$\frac{\sin\sqrt{K}t}{\sqrt{K}}=\frac{2\tan\sqrt{K}t/2}{\sqrt{K}\left(1+\tan^2\sqrt{K}t/2\right)}=\frac{2\,\mathrm{e}^{-t_c}}{1+Ke^{-2t_c}}=\begin{cases}1/\sinh t_c & K=-1\\ 2\,\mathrm{e}^{-t_c} & K=0\\ 1/\cosh t_c & K=+1\end{cases}. \tag{39}$$

With these, (37) becomes

$$\mathrm{d}s^2=\begin{pmatrix}\mathrm{d}t_c^2-\sinh^2 t_c\left(\mathrm{d}r^2+\sinh^2 r\,\mathrm{d}\Omega^2\right) & K=-1\\ \mathrm{d}t_c^2-\mathrm{e}^{2t_c}\,\mathrm{d}\mathbf{x}^2 & K=0\\ \mathrm{d}t_c^2-\cosh^2 t_c\left(\mathrm{d}r^2+\sin^2 r\,\mathrm{d}\Omega^2\right) & K=+1\end{pmatrix}. \tag{40}$$

In the case $K=0$ the freedom of global linear rescaling has been used to make the implicit replacement $r\to 2r$ to cancel the factor of 2 that appears in (39).

## A. Relationship between spacetimes

Eqs. (37) and Eqs. (40) differ only by coordinate transformations, and therefore represent the same spacetime [g)]. The de Sitter spacetime is one of the three maximally symmetric spacetimes that have the same number of positive and negative eigenvalues as Minkowski spacetime and where the Ricci tensor is proportional to the metric tensor [1]. In set notation:

$$\{\text{maximally symmetric}\}=\{\text{de Sitter, Minkowski, anti-de Sitter}\}.$$

Whereas the more general RW spacetimes can be deduced from the Cosmological Principle, the maximally symmetric spacetimes are demanded by the ‘Perfect Cosmological Principle’ of Gold and Bondi [35].

The hyperbolic, flat-space, and spherical RW spatial geometries are not physically distinct because, for example, the de Sitter spacetime is a member of all three (see (35)). In this context it is useful to make a distinction between two different types of coordinate transformation:

---

[g)] These coordinate systems do not cover the manifold equally. The coordinate transformation from $t$ to $t_c$ in (38) is valid only for positive $t$ and therefore the coordinates in (39) do not have the same coverage as the coordinates in, say, (34) [5].

A) *Coordinate transformations that do not mix time and space*: $(t,r)\mapsto(T(t),R(r))$.

Examples are the transformations that take (5) to the forms (6), (7) and (8). In general, such transformations leave preserved the identity of each spatial geometry even though, under a general coordinate transformation, there is some overlap between them. One could say that each spatial geometry is closed under such transformations even if the underlying spacetime, e.g. $\{\text{de Sitter}\}$, is a member of one or more.

B) *Coordinate transformations that mix time and space*: $(t,r)\mapsto(T(t,r),R(t,r))$.

A coordinate transformation of this kind destroys the independence (closure) of the individual spatial geometries. An example is the transformation (24). Another well known example of the overlap between different spatial geometries is the Milne geometry, which is the Minkowski spacetime – usually written $\mathrm{d}s^2=\mathrm{d}x^2$ as a member of $\{\mathrm{RW}_0(t)\}$ (with in this case $a(t)$ fixed) - written instead as a member of $\{\mathrm{RW}_{-1}(t)\}$:

$$\mathrm{d}s^2=\mathrm{d}t^2-t^2\left(\mathrm{d}r^2+\sinh^2 r\,\mathrm{d}\Omega^2\right). \tag{41}$$

The transformation linking this to the form $\mathrm{d}s^2=\mathrm{d}x^2$ involves a transformation of type B that mixes time and space (see for example [16]). Other $\{\mathrm{RW}_{-1}(t)\}$ forms that involve a type A coordinate transformation of (41) are given in Table I.

Noted in passing is that both type A and type B transformations can usefully be split into two sub-types, according to whether or not the functional form of the line element is changed. A trivial example of a type B transformation wherein the line-element appears unchanged is the simple swap $x\leftrightarrow y$ in a Euclidean representation of the Lorentz-Minkowski metric $\mathrm{d}s^2=\mathrm{d}x^2$. A less obvious possibility is the type B transformation

$$T=\frac{t}{t^2-r^2},\quad R=\frac{r}{t^2-r^2}$$

that leaves unchanged the de Sitter line element (34) in polar form,

$$\mathrm{d}s^2=\left(\mathrm{d}t^2-\mathrm{d}r^2-r^2\mathrm{d}\Omega^2\right)/t^2$$

(which very clearly highlights the fact that the coordinates by themselves do not have a distinct physical meaning).

## VI. Conformal form of maximally symmetric spacetimes

Another important restricted class of conformal line elements is associated with $C = \text{constant}$ in (29). After absorbing the factor of ¼ into the coordinates, up to an arbitrary factor the line element is just [1]

$$\mathrm{d}s^2 = \mathrm{d}X^2/\left(1-KX^2\right)^2. \tag{42}$$

(A derivation of the de Sitter case $K$ = 1 was recently given by Lasenby and Doran [36].) Unlike the de Sitter case (34) this line element depends on $K$; here the constraint $C$ is constant does not reduce the three RW spatial geometries to just one spacetime. The (three) line elements (for each $K \in [-1,0,1]$) having the form (8) that transform to (42) can be determined from setting $C$ defined in (30) to a constant (which can be set to 1):

$$\frac{1}{1+Kz_K^2}a^2\left(\frac{1}{\sqrt{K}}\tan^{-1}\sqrt{K}z_K\right)=1 \Rightarrow a^2(t)=1+Kz_K^2=1+\tan^2\sqrt{K}t=\frac{1}{\cos^2\sqrt{K}t}.$$

The corresponding line element in those coordinates is therefore

$$\mathrm{d}s^2=\left(\mathrm{d}t^2-\mathrm{d}r^2-S_K^2(r)\,\mathrm{d}\Omega^2\right)/\cos^2\left(\sqrt{K}t\right). \tag{43}$$

Explicitly:

$$\mathrm{d}s^2=\begin{cases}\left(\mathrm{d}t^2-\mathrm{d}r^2-\sinh^2 r\,\mathrm{d}\Omega^2\right)/\cosh^2 t & K=-1 \quad \text{anti-de Sitter}\\ \mathrm{d}x^2 & K=0 \quad \text{Minkowski}\\ \left(\mathrm{d}t^2-\mathrm{d}r^2-\sin^2 r\,\mathrm{d}\Omega^2\right)/\cos^2 t & K=+1 \quad \text{de Sitter}\end{cases}. \tag{44}$$

The three cases $K \in [-1,0,1]$ therefore pick out the three maximally symmetric spacetimes. Note that the last of (44) is the same as in (37) with the time offset by $\pi/2$. In the extended notation introduced in (32) one can write

$$\{\text{maximally symmetric}\}=\left\{\text{conformal}\,\frac{1}{1-Kx^2}\right\}.$$

Eqs. (43) can be cast into the form (2) with the transformation

$$t_c=\int \mathrm{d}t\,\frac{1}{\cos\sqrt{K}t}=\frac{1}{\sqrt{K}}\log\left(\frac{1+\tan\sqrt{K}t/2}{1-\tan\sqrt{K}t/2}\right)$$

$$\Rightarrow \frac{1+\tan\sqrt{K}t/2}{1-\tan\sqrt{K}t/2}=\mathrm{e}^{\sqrt{K}t_c} \Rightarrow \tan\sqrt{K}t/2=\frac{\mathrm{e}^{\sqrt{K}t_c}-1}{\mathrm{e}^{\sqrt{K}t_c}+1}=\tanh\sqrt{K}t_c/2$$

and therefore

$$\cos^{-2}\left(\sqrt{K}t\right)=\left(\frac{1+\tan^2\left(\sqrt{K}t/2\right)}{1-\tan^2\left(\sqrt{K}t/2\right)}\right)^2=\left(\frac{1+\tanh^2\sqrt{K}t_c/2}{1-\tanh^2\sqrt{K}t_c/2}\right)^2=\cosh^2\left(\sqrt{K}t_c\right).$$

With this, (44) is

$$\mathrm{d}s^2=\mathrm{d}t_c^2-\cosh^2\left(\sqrt{K}t_c\right)\left(\mathrm{d}r^2+S_K^2\left(r\right)\mathrm{d}\Omega^2\right).$$

Explicitly:

$$\mathrm{d}s^2=\begin{cases}\mathrm{d}t_c^2-\cos^2 t_c\left(\mathrm{d}r^2+\sinh^2 r\,\mathrm{d}\Omega^2\right) & K=-1 \quad \text{anti-de Sitter}\\ \mathrm{d}x^2 & K=0 \quad \text{Minkowski}\\ \mathrm{d}t_c^2-\cosh^2 t_c\left(\mathrm{d}r^2+\sin^2 r\,\mathrm{d}\Omega^2\right) & K=+1 \quad \text{de Sitter}\end{cases}. \tag{45}$$

Note that the last of these is the same as the last of (40).

## VII. Conformal form of $\left\{\mathrm{RW}_0\left(t\right)\right\}$ spacetimes

According to (29) and (10) the RW line-element can be written

$$\mathrm{d}s^2=C^2\left(z_K;K\right)\frac{\mathrm{d}X^2}{\left(1-KX^2/4\right)^2} \tag{46}$$

where $C$ is an arbitrary function and $z_K$ is given by (31). When $C=\text{constant}$ the three cases $K\in\left[-1,0,1\right]$ generate the three maximally symmetric spacetimes. However these are not the only possibilities for a conformal factor that is just function of $X$ only. The form of (46) suggests that

$$\mathrm{d}s^2=\mathrm{d}X^2/X^4 \tag{47}$$

may also be a solution corresponding perhaps to some limit involving $K$. In fact (47) is a special case of the conformal line element that is obscured by the normalization procedure in going from (22) to (24) which hampers exploration of limiting behaviors of the solution for particular values of factor $\kappa$ and offsets $\chi_1,\chi_2$.

One possibility is that $\kappa$ is sufficiently small and $\chi_1,\chi_2$ have no special values, with the consequence that only the linear terms in an expansion of the hyperbolic functions in (22) need be retained. In the end the result must be a simple linear transformation of variables taking the RW line element (9) into conformal form. Clearly this can occur only if (9) is restricted to flat space, and indeed it is easy to see from (20) that if $\sigma\left(w\right)$ is linear, then $s^2\left(r\right)\propto r^2$. In the context of the effort of this paper, this linear limit can be ignored since does not give a genuine transformation.

The normalization procedure hides another more interesting limit, however. If one writes $\chi_j = \omega_j + \mathrm{i}\pi/2$ for $j = [1,2]$ then (22) becomes

$$\begin{aligned} T &= B + C + L\left(\coth\left(\kappa(t+r) - \omega_1\right) + \coth\left(\kappa(t-r) - \omega_2\right)\right) \\ \varepsilon R &= B - C + L\left(\coth\left(\kappa(t+r) - \omega_1\right) - \coth\left(\kappa(t-r) - \omega_2\right)\right) \end{aligned}.$$

For sufficiently small values of $\kappa,$ and provided now $\omega_1, \omega_2$ have no special values, then after normalization and removal of offsets one has

$$T = t/x^2, \quad R = r/x^2 \tag{48}$$

- provided $\varepsilon = -1$. Direct substitution of (48) into (11) then gives $s^2(r) = r^2$ and

$$A^2(T,R) = x^4 a^2(t).$$

Eq. (48) is easily inverted to give

$$t = T/X^2, \quad r = R/X^2$$

and therefore the conformal scale factor is related to the flat-space RW scale factor by

$$A^2(T,R) = \frac{a^2\left(T/X^2\right)}{X^4}. \tag{49}$$

Thus the solution (47) is confirmed, corresponding to constant $a(t)$ and signifying Minkowski spacetime. More generally, for arbitrary $a(t)$, the line elements with conformal factor (49) are identical with $\left\{\mathrm{RW}_0(t)\right\}$. In the notation introduced in (32) the correspondence is

$$\left\{\mathrm{RW}_0(t)\right\} = \left\{\mathrm{conformal}\,\frac{1}{x^4}\left(\frac{t}{x^2}\right)\right\} = \left\{\mathrm{conformal}\,\frac{1}{t^2}\left(\frac{t}{x^2}\right)\right\}. \tag{50}$$

## VIII. Conformal form of $\left\{\mathrm{RW}_{-1}(t)\right\}$ spacetimes

This section resumes the investigation of the relationship between the OTI spacetimes (9) and the conformal spacetimes (10). In addition to the hyperbolic transformation discussed in the previous sections, the governing functional equation linking these two spacetimes, (19), has another – exponential – solution, the derivation of which is given in the appendix. It is pointed out there that this exponential solution can also be obtained directly from the hyperbolic solution in the limit that several of the constants have infinite magnitude. It will be seen however that the new transformation permits expression only of the $K = -1$ (hyperbolic space) forms of the RW metrics in conformal form. The limiting procedure therefore implies a change in geometry and is best treated separately.

The appendix results are summarized in (A6). It will be convenient to rewrite those as

$$f(u) = B + \mathrm{e}^{\kappa u + \chi_1 + \chi_2}, \quad g(v) = B + \mathrm{e}^{\kappa v + \chi_1 - \chi_2} \tag{51}$$

so that the transformation is

$$T = 2B + 2\mathrm{e}^{\kappa t + \chi_1} \cosh(\kappa r + \chi_2), \quad \varepsilon R = 2\mathrm{e}^{\kappa t + \chi_1} \sinh(\kappa r + \chi_2) \tag{52}$$

where $B, \kappa, \chi_1, \chi_2$ are constants. Inserting (51) into (16) gives

$$s^2(r) = \frac{\left(e^{\kappa u + \chi_1 + \chi_2} - e^{\kappa v + \chi_1 - \chi_2}\right)^2}{4\kappa^2 e^{\kappa(u+v) + 2\chi_1}} = \frac{1}{\kappa^2} \sinh^2(\kappa r + \chi_2)\,. \tag{53}$$

Remarkably, the only allowed spatial curvature functions in the OTI spacetimes compatible with the exponential transformation to conformal form are those compatible with the RW spacetimes, just as in the case of the hyperbolic solution. Unlike the hyperbolic transformations however, all three RW spatial geometries cannot be accommodated with this solution to the functional equation (19). For the transformation (52) to be real and not vanishing it is clear that $\kappa$ must be real and non-zero. Therefore only the subset $\{\mathrm{RW}_{-1}(t)\}$ can be transformed to conformal form with (51) [h]. Rescaling the variables and allowing for implicit offsets, (51) becomes

$$f(u) = \mathrm{e}^u, \quad g(v) = \mathrm{e}^v\,, \tag{54}$$

(52) becomes

$$T = \mathrm{e}^t \cosh r, \quad R = \mathrm{e}^t \sinh r\,, \tag{55}$$

and (53) becomes

$$s^2(r) = \sinh^2 r\,.$$

As anticipated in (25), combining this finding with that of Sec. IV establishes the general result

---

[h] In an appropriate limit, wherein $\kappa \to 0$ and $\chi_1 = -\log\kappa$, the transformation (51) becomes linear, and up to a simple rescaling (52) becomes the trivial (‘identity’) transformation. In that sense only, (51) also encompasses the $\{\mathrm{RW}_0(t)\}$ metrics, trivially transforming them onto themselves.

$$\left\{\mathrm{RW}_K\left(t\right)\right\}=\left\{\mathrm{OTI}\left(t\right)\left(r\right)\right\}\cap\left\{\mathrm{conformal}\left(t,r\right)\right\},$$

which relation is expressed in Fig. I.

The conformal scale factor associated with the exponential transformation can be obtained from (26) wherein it may be recalled that $a^2\left(t\right)$ is the RW scale factor of (9f). Using (54) in (26) gives

$$A^2\left(T,R\right)=\frac{1}{4}a^2\left(t\right)\mathrm{e}^{-2t}\,. \tag{56}$$

Eq. (55) is easily inverted. Specifically, one has

$$X^2\equiv T^2-R^2=\mathrm{e}^{2t} \tag{57}$$

which establishes that the transformation is valid only inside the light-cone in the ($T$,$R$) coordinate system. Eq. (56) with (57) is

$$A^2\left(T,R\right)=\frac{1}{4X^2}a^2\left(\frac{1}{2}\log X^2\right). \tag{58}$$

Since the RW scale factor is arbitrary, it follows that the conformal scale factor is an arbitrary function of $X^2$.

In summary, the hyperbolic RW metric with arbitrary scale factor $a^2\left(t\right)$ can be transformed with (55) to conformal form with scale factor given by (58). In the domains where both sets of coordinates are real the two spacetimes are therefore the same. Further, it may be observed that (58) includes the more restricted forms of scale factors given in (42). Given the subsequent classification for example in (44), it follows that all three maximally symmetric spacetimes may be cast into the hyperbolic form of the RW metric - as may be verified by examination of the entries in Table I. It follows that

$$\left\{\mathrm{RW}_{-1}\left(t\right)\right\}=\left\{\mathrm{conformal}\left(x^2\right)\right\}\supset\left\{\text{maximally symmetric}\right\}. \tag{59}$$

It is important to observe that the conformal factor in (58) is functionally quite different and cannot, in general, be obtained from (29) when $K=-1$ by any redefinition of the functional degrees of freedom. This is despite the fact that, prior to choosing the constant degrees of freedom, they both derive from the same general solution (A23) to the functional equation (17) (see Section A4).

Finally, using the results (58) and (49) it is possible to show that

$$\left\{\mathrm{RW}_{-1}(t)\right\} \cap \left\{\mathrm{RW}_{0}(t)\right\} \setminus \left\{\mathrm{RW}_{+1}(t)\right\} = \left\{\text{Minkowski}\right\}$$

which, bearing in mind (35), gives

$$\left\{\mathrm{RW}_{-1}(t)\right\} \cap \left\{\mathrm{RW}_{0}(t)\right\} = \left\{\text{Minkowski}\right\} \cup \left\{\text{de Sitter}\right\}.$$

The $\mathrm{RW}_{-1}(t)$ form here is the Milne metric. The common ground of the RW metrics is therefore exclusively in the maximally symmetric spacetimes, with de Sitter being uniquely expressible in all three $K \in [-1, 0, 1]$, Minkowski expressible in $K \in [-1, 0]$, and anti-de Sitter expressible only in $K = -1$. These relationships are depicted in Fig. I.

## IX. Summary

Eq. (32) gives the conformal representation of all the RW spacetimes, and Eq. (59) gives a conformal representation of the RW hyperbolic space spacetime. We give the general solution (A20) to the functional equation (A1). And we give the general solution (A23) and a particular solution (A25) – independently and as a limiting case of (A23) - to the functional equation (17). Historically, the general solution (A25) has been used to generate the hyperbolic transformations and associated conformal factors (29), and except in a few cases, the exponential transformation and associated (different) conformal factor (58) have been missed.

An outcome of this analysis is the catalogue of line elements in Table I, together with coordinate transformations linking the different forms in the body of this document. Table I includes, in line-element form, an exhaustive list of all possible cosmological (RW-compliant) metrics that are conformal to the Lorentz-Minkowski metric. The table is not an exhaustive list of RW metrics however, even within the restricted domain of diagonal forms. Not included for example are diagonal harmonic forms of the RW spacetimes [37]. Non-diagonal forms, for example as comprehensively studied by Weinberg [1], have not been considered. A product of this approach is the determination of set-theoretical relationships between the cosmologically-relevant spacetimes, and which are summarized in Fig. I.

## X. Acknowledgements

The author is very grateful to Anthony Lasenby, János Aczél and the referee for help and advice, and to Sergei Ketov for alerting me to his own work. Special thanks to Aidan Keene for providing important missing references, including background to the conformal symmetry.

**Table I.**

Catalogue of cosmologically-significant diagonal line elements discussed in this document.

$\{\text{conformal}(t,r)\}$

| line element | | text references | |
|---|---|---|---|
| $f^2(t,r)\mathrm{d}x^2$ | | (10) | |

$\{\text{OTI}(t)(r)\}$

| line element | | text references | |
|---|---|---|---|
| $\mathrm{d}t^2 - a^2(t)s^2(r)\left(\mathrm{d}r^2 + r^2\mathrm{d}\Omega^2\right)$ | | (9a) | |
| $a^2(t)\left(\mathrm{d}t^2 - s^2(r)\left(\mathrm{d}r^2 + r^2\mathrm{d}\Omega^2\right)\right)$ | | (9b) | |
| $\mathrm{d}t^2 - a^2(t)\left(\mathrm{d}r^2 + s^2(r)\mathrm{d}\Omega^2\right)$ | | (9c) | |
| $a^2(t)\left(\mathrm{d}t^2 - \mathrm{d}r^2 - s^2(r)\mathrm{d}\Omega^2\right)$ | | (9d) | |
| $\mathrm{d}t^2 - a^2(t)\left(\mathrm{d}r^2 + s^2(r)\mathrm{d}\Omega^2\right)$ | | (9e) | |
| $a^2(t)\left(\mathrm{d}t^2 - \mathrm{d}r^2 - s^2(r)\mathrm{d}\Omega^2\right)$ | | (9f) | |

$\{\text{RW}_K(t)\}$

| line element | coordinate system | text references | literature references |
|---|---|---|---|
| $dt^2 - f^2(t)\mathrm{d}\Sigma_K^2$ | RW | (5),(6),(7) | common use |
| $f^2(t)\left(\mathrm{d}t^2 - \mathrm{d}\Sigma_K^2\right)$ | conformal to static RW | (8) | |
| $\frac{1}{t^2} f^2\left(\frac{t}{1-Kx^2}\right)\mathrm{d}x^2$ | conformal to Lorentz-Minkowski | (29) | 21 |
| $\frac{1}{\left(1-Kx^2\right)^2} f^2\left(\frac{t}{1-Kx^2}\right)\mathrm{d}x^2$ | conformal to Lorentz-Minkowski | (29) | |

$\{\mathrm{RW}_{-1}(t)\}$

| line element | coordinate system | text references | literature references |
|---|---|---|---|
| $\mathrm{d}t^2 - f^2(t)\mathrm{d}\Sigma_{-1}^2$ | RW K=-1 | (5),(6),(7) | common use |
| $f^2(t)\left(\mathrm{d}t^2 - \mathrm{d}\Sigma_{-1}^2\right)$ | conformal to static RW K=-1 | (8) | |
| $\frac{1}{t^2} f^2\left(\frac{t}{1+x^2}\right)\mathrm{d}x^2$ | conformal to Lorentz-Minkowski | (29) | 21,36 |
| $\frac{1}{\left(1+x^2\right)^2} f^2\left(\frac{t}{1+x^2}\right)\mathrm{d}x^2$ | conformal to Lorentz-Minkowski | (29) | |
| $f^2\left(x^2\right)\mathrm{d}x^2$ | conformal to Lorentz-Minkowski | (58) | 21,38 |

$\{\mathrm{RW}_{0}(t)\}$

| line element | coordinate system | text references | literature references |
|---|---|---|---|
| $\mathrm{d}t^2 - f^2(t)\mathrm{d}\Sigma_{0}^2$ | RW K=0 | (5),(6),(7) | common use |
| $f^2(t)\left(\mathrm{d}t^2 - \mathrm{d}\Sigma_{0}^2\right)$ | conformal to Lorentz-Minkowski | (8) | |
| $f^2\left(t/x^2\right)\mathrm{d}x^2/t^2$ | conformal to Lorentz-Minkowski | (50) | 21 |
| $f^2\left(t/x^2\right)\mathrm{d}x^2/x^4$ | conformal to Lorentz-Minkowski | (50) | |

$\{\mathrm{RW}_{+1}(t)\}$

| line element | coordinate system | text references | literature references |
|---|---|---|---|
| $\mathrm{d}t^2 - f^2(t)\mathrm{d}\Sigma_{+1}^2$ | RW K=+1 | (5),(6),(7) | common use |
| $f^2(t)\left(\mathrm{d}t^2 - \mathrm{d}\Sigma_{+1}^2\right)$ | conformal to static RW K=+1 (Einstein) | (8) | |
| $\frac{1}{t^2} f^2\left(\frac{t}{1-x^2}\right)\mathrm{d}x^2$ | conformal to Lorentz-Minkowski | (29) | 21,36 |
| $\frac{1}{\left(1-x^2\right)^2} f^2\left(\frac{t}{1-x^2}\right)\mathrm{d}x^2$ | conformal to Lorentz-Minkowski | (29) | |

{maximally symmetric}

| line element | coordinate system | text references | literature references |
|---|---|---|---|
| $dt^2 - \cosh^2\left(\sqrt{K}t\right) d\Sigma_K^2$ | RW | (45) | |
| $\left(dt^2 - d\Sigma_K^2\right)/\cos^2\left(\sqrt{K}t\right)$ | conformal to static RW | (43) | |
| $dx^2/\left(1-Kx^2\right)^2$ | conformal to Lorentz-Minkowski | (42) | 1,12 |
| $dx^2/\left(K-x^2\right)^2$ | conformal to Lorentz-Minkowski | (42) & (47) | |

{anti-de Sitter}

| line element | coordinate system | text references | literature references |
|---|---|---|---|
| $dt^2 - \cos^2 t\, d\Sigma_{-1}^2$ | RW K=-1 | (45) | 5 |
| $\left(dt^2 - d\Sigma_{-1}^2\right)/\cosh^2 t$ | conformal to static RW K=-1 | (44) | |
| $dx^2/\left(1+x^2\right)^2$ | conformal to Lorentz-Minkowski | (42) | 39 |

{Minkowski}

| line element | coordinate system | text references | literature references |
|---|---|---|---|
| $dx^2$ | Lorentz | | usual form |
| $dt^2 - t^2\, d\Sigma_{-1}^2$ | Milne | (41) | 21,40,41 |
| $e^{2t}\left(dt^2 - d\Sigma_{-1}^2\right)$ | conformal to static RW K=-1 | | |
| $dx^2/x^4$ | conformal to Lorentz-Minkowski | (47) | 21 † |

$\{\text{de Sitter}\}$

| line element | coordinate system | text references | literature references |
|---|---|---|---|
| $\mathrm{d}t^2 - \sinh^2 t\,\mathrm{d}\Sigma_{-1}^2$ | RW K=-1 | (40) | 8 |
| $\left(\mathrm{d}t^2 - \mathrm{d}\Sigma_{-1}^2\right)/\sinh^2 t$ | conformal to static RW K=-1 | (37) | |
| $\mathrm{d}t^2 - \mathrm{e}^{2t}\,\mathrm{d}\Sigma_0^2$ | RW K=0 | (40) | 1,2,6,8,12,21 |
| $\mathrm{d}x^2 / t^2$ | conformal to Lorentz-Minkowski | (37) | |
| $\mathrm{d}t^2 - \cosh^2 t\,\mathrm{d}\Sigma_1^2$ | RW K=1 | (40) | 2,8,12 |
| $\left(\mathrm{d}t^2 - \mathrm{d}\Sigma_1^2\right)/\cos^2 t$ | conformal to static RW K=+1 | (37) | |
| $\mathrm{d}x^2/\left(1-x^2\right)^2$ | conformal to Lorentz-Minkowski | (42) | 21,36,39,42 |

† The published metric is in error.

**Notes on the table entries**

- For each $K$, $\mathrm{d}\Sigma_K^2$ is any of the equivalent forms

$$\mathrm{d}\Sigma_1^2 = \begin{cases} \mathrm{d}r^2 + \sin^2 r\,\mathrm{d}\Omega^2 \\ \dfrac{\mathrm{d}r^2}{1-r^2} + r^2\mathrm{d}\Omega^2 \\ \left(\mathrm{d}r^2 + r^2\mathrm{d}\Omega^2\right)/\left(1+r^2/4\right)^2 \end{cases}, \quad \mathrm{d}\Sigma_0^2 = \begin{cases} \mathrm{d}\mathbf{x}^2 \\ \mathrm{d}r^2 + r^2\mathrm{d}\Omega^2 \end{cases}, \quad \mathrm{d}\Sigma_{-1}^2 = \begin{cases} \mathrm{d}r^2 + \sinh^2 r\,\mathrm{d}\Omega^2 \\ \dfrac{\mathrm{d}r^2}{1+r^2} + r^2\mathrm{d}\Omega^2 \\ \left(\mathrm{d}r^2 + r^2\mathrm{d}\Omega^2\right)/\left(1-r^2/4\right)^2 \end{cases}$$

  or any other equivalent spatial metric obtained by a transformation solely of the spatial coordinates.

- $$x^2 \equiv t^2 - \mathbf{x}^2, \quad \mathrm{d}x^2 \equiv \mathrm{d}t^2 - \mathrm{d}\Sigma_0^2$$

- It is understood that every line element can be scaled by an arbitrary constant (conformal) scale factor without changing the classification of the spacetime. Further, all coordinates are understood to be normalized with respect to some fixed length (which is in principle observable). That is, for a general diagonal line element

$$\mathrm{d}s^2 = g_{ij}\left(x\right)dx^i dx^j\,.$$

  The classification is unchanged if

$$\mathrm{d}s^2 \to \alpha\beta_i^{\ m}\beta_j^{\ n} g_{mn}\left(\beta x\right)dx^i dx^j$$

where $\beta$ is a diagonal 4x4 matrix and $\alpha$ is an additional arbitrary constant. (The arbitrary rescaling of the interval $\mathrm{d}s$ is not generally the same as a coordinate transformation.) It follows that the trigonometric functions sine and cosine can be interchanged so that $\sin\left(\sqrt{K}t\right)/\sqrt{K}$ can be replaced by $\cos\left(\sqrt{K}t\right)/\sqrt{K}$ if $K$ is a real positive number but not otherwise.

- Everywhere $f\left(z\right)$ is an arbitrary function, each occurrence of which is to be treated de novo.
- Wherever it appears, $K \in \left[-1,0,1\right]$.

## Appendix A: General solution of the functional equation

$$f(u)-g(v)=\phi(u)\gamma(v)\sigma(u-v). \tag{A1}$$

A general method for the solution of a class of functional equations of which (A1) is a member is given in Aczél [43], although this particular equation is not solved explicitly there. Here a standard technique is employed for solution of multivariate functional equations wherein all but one functional dependency is eliminated by differentiation, the resulting differential equation is solved, and then the results are reconciled with the original functional equation. For the sake of completeness, in this appendix Eq. (A1) is solved for all five functions. However, the pair $\phi(u),\gamma(v)$ play no role in the subsequent analysis in the body of this paper.

Differentiating (A1) with respect to both $u$ and $v$ and using $\phi=\mathrm{e}^{p},\gamma=\mathrm{e}^{q}$ gives

$$\phi'\gamma'\sigma+\sigma'\left(\phi\gamma'-\gamma\phi'\right)-\phi\gamma\sigma''=0\Rightarrow -p'q'+\frac{\sigma'}{\sigma}\left(p'-q'\right)+\frac{\sigma''}{\sigma}=0 \tag{A2}$$

provided $\phi,\gamma,\sigma\neq 0$. $\sigma$ can be eliminated by differentiation with respect to t, i.e. by $\partial/\partial u+\partial/\partial v$:

$$-p''q'-p'q''+\frac{\sigma'}{\sigma}\left(p''-q''\right)=0. \tag{A3}$$

### A1. Constant case

The case that the second derivative of just one of the two functions $p$ and $q$ vanishes can be discounted as follows. If say $p''=0,\quad q''\neq 0$ then $p=a+bu$ for constant $a$ and $b$, and (A3) then gives

$$b+\sigma'/\sigma=0\Rightarrow\sigma=\sigma_0\,\mathrm{e}^{-b(u-v)}$$

where $\sigma_0$ is constant. Putting these results into (A1) gives

$$f(u)-g(v)=\sigma_0\,\mathrm{e}^{a+bv}\,\gamma(v).$$

Clearly $f(u)$ must be a constant. Swapping the roles of $p$ and $q$ gives that $g(v)$ is constant. Neither of these two possibilities are acceptable solutions in this context, since they imply a loss of dimensionality in the transformation.

### A2. Exponential case

A particular solution of (A3) is $p''=q''=0$ whereupon $p$ and $q$ are linear in their arguments:

$$p(u) = \alpha_1 + \beta_1 u, \quad q(v) = \alpha_2 + \beta_2 v \tag{A4}$$

where the $\alpha_i, \beta_i$ are constants. Recalling (A2) this implies

$$\sigma'' + \sigma'(\beta_1 - \beta_2) - \beta_1\beta_2\sigma = 0 \Rightarrow \sigma = \alpha_3\, e^{-\beta_1(u-v)} + \alpha_4\, e^{\beta_2(u-v)} \tag{A5}$$

provided both $\beta_i$ are not zero. $\alpha_3, \alpha_4$ are new constants. Putting these into (A1) gives

$$f(u) - g(v) = e^{\alpha_1+\alpha_2+\beta_1 u+\beta_2 v}\left(\alpha_3\, e^{-\beta_1(u-v)} + \alpha_4\, e^{\beta_2(u-v)}\right).$$

It follows that one solution of (A1) is the set of functions

$$\begin{aligned} \phi(u) &= a_1\, e^{\beta_1 u}, \quad \gamma(v) = a_2\, e^{\beta_2 v} \\ \sigma(u-v) &= a_3\, e^{-\beta_1(u-v)} + a_4\, e^{\beta_2(u-v)} \\ f(u) &= a_5 + a_1 a_2 a_4\, e^{(\beta_1+\beta_2)u} \\ g(v) &= a_5 - a_1 a_2 a_3\, e^{(\beta_1+\beta_2)v} \end{aligned} \tag{A6}$$

where $a_i, \beta_i$ are constants.

## A2.1 Linear case

If $\beta_1 = \beta_2 = 0$ in (A4) then $p$ and $q$ are constant,

$$p = \alpha_1, \quad q = \alpha_2,$$

in which case instead of (A5) one has

$$\sigma'' = 0 \Rightarrow \sigma = \alpha_3 + \beta(u-v).$$

Putting these into (A1) gives

$$f(u) - g(v) = e^{\alpha_1+\alpha_2}\left(\alpha_3 + \beta(u-v)\right).$$

It follows that one solution of (A1) is the set of functions

$$\begin{aligned} \phi(u) &= a_1, \quad \gamma(v) = a_2 \\ \sigma(u-v) &= a_3 + \beta(u-v) \\ f(u) &= a_4 + a_1 a_2 \alpha_3 / 2 + a_1 a_2 \beta u \\ g(v) &= a_4 - a_1 a_2 \alpha_3 / 2 + a_1 a_2 \beta v \end{aligned}. \tag{A7}$$

This result can be obtained from (A6) by taking suitable limiting values of the constants.

## A3. Hyperbolic case

Here is considered the case that $p''$ and $q''$ are both non-zero, in which case it is possible to rearrange (A3) and then operate again with $\partial/\partial u + \partial/\partial v$ to remove the dependency on $\sigma$:

$$\begin{aligned}&\frac{\sigma'}{\sigma} = \frac{p''q' + p'q''}{p'' - q''}\\ &\Rightarrow \frac{p'''q' + 2p''q'' + p'q'''}{p'' - q''} - \frac{\left(p''q' + p'q''\right)\left(p''' - q'''\right)}{\left(p'' - q''\right)^2} = 0\\ &\Rightarrow \left(p'''q' + 2p''q'' + p'q'''\right)\left(p'' - q''\right) - \left(p''q' + p'q''\right)\left(p''' - q'''\right) = 0\end{aligned} .$$

This can be rearranged to give

$$\left(2p'' - p'\frac{p'''}{p''}\right) + \left(q'\frac{q'''}{q''} - 2q''\right) - q'\frac{p'''}{p''} + p'\frac{q'''}{q''} = 0 . \qquad \text{(A8)}$$

The first term in parentheses is a function of $u$ only, and the second term a function of $v$ only. Therefore differentiation with respect to both $u$ and $v$ causes them both to vanish, whereupon

$$q''\left(\frac{p'''}{p''}\right)' - p''\left(\frac{q'''}{q''}\right)' = 0 . \qquad \text{(A9)}$$

Separation of the $u$ and $v$ dependent parts then gives

$$\frac{\lambda}{2}\left(\frac{p'''}{p''}\right)' = p''$$

and likewise for $q$. $\lambda$ is a constant of separation. Writing $p'' = 1/\psi^2$ this is

$$\psi'^2 - \psi\psi'' = 1/\lambda . \qquad \text{(A10)}$$

Differentiating again

$$\psi'\psi'' = \psi\psi''' \Rightarrow \frac{\mathrm{d}}{\mathrm{d}u}\log\psi = \frac{\mathrm{d}}{\mathrm{d}u}\log\psi'' \Rightarrow \psi'' = k^2\psi$$

for any (possibly complex) $k$. The general solution is

$$\psi = a\,\mathrm{e}^{\kappa u} + b\,\mathrm{e}^{-\kappa u} \qquad \text{(A11)}$$

where $a, b, \kappa$ are constants. The constants are not independent however. Substitution of (A11) into (A10) gives the relation

$$\left(ae^{\kappa u} - be^{-\kappa u}\right)^2 - \left(ae^{\kappa u} + be^{-\kappa u}\right)^2 + 1/\lambda\kappa^2 = 0 \Rightarrow 4\lambda\kappa^2 ab = 1 . \qquad \text{(A12)}$$

Putting (A11) into (A9) gives

$$p=\int^{u}\mathrm{d}u'\int^{u'}\mathrm{d}u''\frac{1}{\left(a_1\,\mathrm{e}^{\kappa_1 u''}+b_1\,\mathrm{e}^{-\kappa_1 u''}\right)^2}=-\frac{1}{4\kappa_1^2 a_1 b_1}\left(2\kappa_1 u+\log\left(a_1\,\mathrm{e}^{2\kappa_1 u}+b_1\right)\right)+c_1+d_1 u\,. \qquad \text{(A13)}$$

Using (A12) and redefining the constants (A13) and the corresponding equation for $q$ can be written

$$\begin{aligned} p(u)&=\alpha_1+\beta_1 u-\lambda\log\left(\gamma_1+\mathrm{e}^{2\kappa_1 u}\right)\\ q(v)&=\alpha_2+\beta_2 v-\lambda\log\left(\gamma_2+\mathrm{e}^{2\kappa_2 v}\right) \end{aligned}. \qquad \text{(A14)}$$

These constants are still not independent. Substitution of these expressions into (A8) gives the relations

$$\begin{aligned} \gamma_1\gamma_2(\kappa_1-\kappa_2)(\beta_1+\beta_2)&=0\\ \gamma_2(\kappa_1+\kappa_2)(\beta_1+\beta_2+2\lambda\kappa_1)&=0\\ \gamma_1(\kappa_1+\kappa_2)(\beta_1+\beta_2+2\lambda\kappa_2)&=0\\ (\kappa_1-\kappa_2)\left(\beta_1+\beta_2+2\lambda(\kappa_1+\kappa_2)\right)&=0 \end{aligned}. \qquad \text{(A15)}$$

The possibility $\lambda = 0$ converts (A14) to (A4) and can therefore be excluded here. The remaining cases are given in Table AI.

**Table AI.**

List of constraints satisfying (A15).

| Case | Constraints | Comments |
|---|---|---|
| I | $\kappa_1 = \kappa_2 = \kappa, \quad \beta_1 + \beta_2 + 2\lambda\kappa = 0$ | genuinely new case; discussed below |
| II | $\beta_1 = -\beta_2, \quad \kappa_1 = -\kappa_2$ | re-presentation of Case I – see below |
| III | $\kappa_1 = \kappa_2 = 0$ | reduces (A14) to (A4) |
| IV | $\gamma_1 = \gamma_2 = 0, \quad \beta_1 + \beta_2 + 2\lambda\left(\kappa_1 + \kappa_2\right) = 0$ | reduces (A14) to (A4) |
| V | $\gamma_1 = \gamma_2 = 0, \quad \kappa_1 = \kappa_2$ | reduces (A14) to (A4) |
| VI | $\gamma_1 = 0, \quad \kappa_2 = 0, \quad \beta_1 + \beta_2 + 2\lambda\kappa_1 = 0$ | reduces (A14) to (A4) |
| VII | $\gamma_2 = 0, \quad \kappa_1 = 0, \quad \beta_1 + \beta_2 + 2\lambda\kappa_2 = 0$ | reduces (A14) to (A4) |
| VIII | $\gamma_1 = 0, \quad \kappa_1 = \kappa_2 = \kappa, \quad \beta_1 + \beta_2 + 2\lambda\kappa = 0$ | special case of I |
| IX | $\gamma_1 = 0, \quad \kappa_1 = -\kappa_2, \quad \beta_1 = -\beta_2$ | special case of II |
| X | $\gamma_2 = 0, \quad \kappa_1 = \kappa_2 = \kappa, \quad \beta_1 + \beta_2 + 2\lambda\kappa = 0$ | special case of I |
| XI | $\gamma_2 = 0, \quad \kappa_1 = -\kappa_2, \quad \beta_1 = -\beta_2$ | special case of II |

### A3.1 Case I

Using

$$\kappa_1 = \kappa_2 = \kappa$$
$$\beta_1 + \beta_2 = 2\lambda\kappa, \quad \beta_1 - \beta_2 = 2\beta_\Delta$$
$$\Rightarrow \beta_1 = \beta_\Delta - \lambda\kappa, \quad \beta_2 = -\beta_\Delta - \lambda\kappa$$

Eq. (A14) becomes

$$\begin{aligned} p(u) &= \alpha_1 + (\beta_\Delta - \lambda\kappa)u - \lambda \log\left(\gamma_1 + \mathrm{e}^{-2\kappa u}\right) \\ q(v) &= \alpha_2 - (\beta_\Delta + \lambda\kappa)v - \lambda \log\left(\gamma_2 + \mathrm{e}^{-2\kappa v}\right) \end{aligned}.$$

With the substitutions

$$\gamma_1 = e^{2\chi_1}, \quad \gamma_2 = e^{2\chi_2}$$

these expressions can be put into the form

$$\begin{aligned} p(u) &\to \alpha_1 + \beta_\Delta u - \lambda \log\cosh(\kappa u + \chi_1) \\ q(v) &\to \alpha_2 - \beta_\Delta v - \lambda \log\cosh(\kappa v + \chi_2) \end{aligned} \tag{A16}$$

with the $\alpha_i$ suitably redefined, and therefore

$$\phi(u) = a_1 \frac{\mathrm{e}^{\beta_\Delta u}}{\cosh^\lambda(\kappa u + \chi_1)}, \quad \gamma(v) = a_2 \frac{\mathrm{e}^{-\beta_\Delta v}}{\cosh^\lambda(\kappa v + \chi_2)}. \tag{A17}$$

Inserting (A16) into (A3) gives

$$\frac{\sigma'}{\sigma} = \beta_\Delta - \lambda\kappa \coth\left(\kappa(u - v) + \chi_1 - \chi_2\right).$$

Integrating:

$$\sigma = a_3 \exp\left(-\beta_\Delta\left(u - v + (\chi_1 - \chi_2)/\kappa\right)\right)\sinh^\lambda\left(\kappa(u - v) + \chi_1 - \chi_2\right).$$

Putting this and (A17) into (A1) gives

$$f(u) - g(v) = a_1 a_2 a_3\, \mathrm{e}^{-\beta_\Delta(\chi_1 - \chi_2)/\kappa}\left[\frac{\sinh\left(\kappa(u - v) + \chi_1 - \chi_2\right)}{\cosh(\kappa u + \chi_1)\cosh(\kappa v + \chi_2)}\right]^\lambda. \tag{A18}$$

Expanding the numerator one obtains

$$f(u) - g(v) = a_1 a_2 a_3 2^{-\lambda} e^{-\beta_\Delta(\chi_1 - \chi_2)/\kappa}\left(\tanh(\kappa u + \chi_1) - \tanh(\kappa v + \chi_2)\right)^\lambda. \tag{A19}$$

Clearly, disregarding the trivial case $\lambda = 0$, the above demands that $\lambda = 1$. Therefore a particular solution of (A1) is the set of functions

$$\begin{aligned} \phi(u) &= a_1 \frac{\mathrm{e}^{\beta u}}{\cosh(\kappa u + \chi_1)}, \quad \gamma(v) = a_2 \frac{\mathrm{e}^{-\beta v}}{\cosh(\kappa v + \chi_2)} \\ \sigma(u-v) &= a_3 e^{-\beta(u-v)} \sinh(\kappa(u-v) + \chi_1 - \chi_2) \\ f(u) &= a_4 + a_1 a_2 a_3 \tanh(\kappa u + \chi_1) \\ g(v) &= a_4 + a_1 a_2 a_3 \tanh(\kappa v + \chi_2) \end{aligned} \tag{A20}$$

where $c_i, \beta, \kappa, \chi_i$ are constants.

## A3.2 Case II

Using

$$\begin{aligned} &\kappa_1 = -\kappa_2 = \kappa \\ &\beta_1 + \beta_2 = 0, \quad \beta_1 - \beta_2 = 2(\beta_\Delta - \lambda\kappa) \, . \\ &\Rightarrow \beta_1 = \beta_\Delta - \lambda\kappa, \quad \beta_2 = -\beta_\Delta + \lambda\kappa \end{aligned} \tag{A21}$$

Eq. (A14) becomes

$$\begin{aligned} p(u) &= \alpha_1 + (\beta_\Delta - \lambda\kappa)u - \lambda \log(\gamma_1 + \mathrm{e}^{-2\kappa u}) \\ q(v) &= \alpha_2 - (\beta_\Delta - \lambda\kappa)v - \lambda \log(\gamma_2 + \mathrm{e}^{2\kappa v}) \end{aligned} . \tag{A22}$$

With suitable redefinition of the constants (A22) is the same as (A16). Since $\sigma(u-v)$ is given entirely in terms of $p(u), q(v)$ (through (A3)), it follows that Case II is indistinguishable from Case I.

## A4. Relationship between solutions

Eq. (A6) can reduce to (A7), and (A7) is equivalent to (A20), so (A7) is not an independent solution of (A1).

Eq. (A20) will reduce to (A6) in a limit in which the magnitudes of $a, b, \chi_1$ and $\chi_2$ are all infinite. Specifically, starting from

$$f(u) = a + b \tanh(\kappa u + \chi_1), \quad g(v) = a + b \tanh(\kappa v + \chi_2), \tag{A23}$$

the substitutions

$$a = \alpha + \frac{\gamma}{2\delta^2}, \quad b = \frac{\gamma}{2\delta^2}, \quad \chi_1 = \chi_2 = \log \delta \tag{A24}$$

give

$$f(u)=\alpha+\beta e^{2\kappa u}+O\left(\delta^2\right),\quad g(v)=\alpha+\gamma e^{2\kappa v}+O\left(\delta^2\right). \tag{A25}$$

Taking the limit that $\delta\to 0$ then gives effectively the same transformations as (A6).

## A5. Connection with the method of Lightman et al

Lightman et al address the solution of (17) without going via the functional equation, (19). They demand at the outset that the two undetermined functions must be equal (which turns out to be correct up to linear scaling and offset) and so set out to solve an equation equivalent to

$$\frac{4\dot{g}(u)\dot{g}(v)}{\left(g(u)-g(v)\right)^2}=\frac{1}{s^2\left((u-v)/2\right)} \tag{26}$$

where, a priori, they are concerned only with the particular case that

$$4s^2\left((u-v)/2\right)=\left\{\sinh(u-v),u-v,\sin(u-v)\right\} \tag{27}$$

corresponding to $K=-1,0,+1$ in the RW metrics. (An outcome of the more general approach taken above is that these are found to be the *only* possibilities for the function *s* solution of (26).) However, Lightman et al solve (27) explicitly only for the particular $K=+1$ from which they subsequently attempt to infer the result for $K=-1$. Below we recap their method whilst retaining $K$ as a variable in order to accommodate both the hyperbolic and exponential solutions, and to help keep track of where the latter goes missing.

Since (17) must be valid for any *u* and *v*, one can write $u=v+\varepsilon$ and require that it be valid for any $\varepsilon$. In particular, expanding for small $\varepsilon$

$$\frac{\dot{g}(u)\dot{g}(u+\varepsilon)}{\left(g(u)-g(u+\varepsilon)\right)^2}=\frac{1}{4s^2(\varepsilon/2)}$$

$$\Rightarrow\frac{\dot{g}(u)\left(\dot{g}(u)+\varepsilon\ddot{g}(u)+\varepsilon^2\dddot{g}(u)/2+O\left(\varepsilon^3\right)\right)}{\left(\varepsilon\dot{g}(u)+\varepsilon^2\ddot{g}(u)/2+\varepsilon^3\dddot{g}(u)/6+O\left(\varepsilon^3\right)\right)^2}=\frac{1}{\left(\varepsilon-K\varepsilon^3/6+O\left(\varepsilon^5\right)\right)^2}$$

Collecting powers of $\varepsilon$ one finds that the first two terms in the expansion are true independent of *g*, and that the first constraint on *g* is

$$2\dot{g}\dddot{g}-3\ddot{g}^2-\frac{4}{K}\dot{g}^2=0. \tag{28}$$

It may be noted that $g\propto e^{\beta u}$ is a solution of (28) if $\beta=\pm 2/\sqrt{-K}$, and therefore provided $K=-1$. However, since Lightman et al carry out this calculation explicitly only for the particular case $K=+1$, the exponential solution was not strictly available to them from this point onwards [i]. With the substitutions $p=\dot{g}, \quad q=\ddot{g} \Rightarrow \dddot{g}=dq/du=qdq/dp$ (28) becomes

$$p\frac{dq^2}{dp}-3q^2-\frac{4}{K}p^2=0\,.$$

Determination of the first integral is facilitated by an integrating factor

$$p^4\frac{d}{dp}\left(\frac{q^2}{p^3}\right)=\frac{4}{K}p^2 \Rightarrow q=p\sqrt{Ap-4/K} \tag{29}$$

where $A$ is a constant. The exponential solution corresponds to the special value $A=0$ (provided $K=-1$). After performing the first integral of (29) to give

$$u+B=\int dp\frac{1}{p\sqrt{Ap-4/K}}=\sqrt{K}\tan^{-1}\sqrt{ApK/4-1} \tag{30}$$

the exponential solution no longer corresponds exactly to $A=0$, but must be found by a limiting procedure in which $A\to 0$ and $|B|\to\infty$ (again, supposing $K=-1$ had been explicitly retained). Finally, solving (30) one obtains

$$\begin{aligned} &ApK/4=1+\tan^2\frac{u+B}{\sqrt{K}} \Rightarrow \frac{dg}{du}=\frac{4}{AK}\sec^2\frac{u+B}{\sqrt{K}} \\ &\Rightarrow g=\frac{4}{A\sqrt{K}}\tan\frac{u+B}{\sqrt{K}}+C \end{aligned}\,. \tag{31}$$

After linear rescaling and offset, Lightman et al give $g=\tan u$ for $K=+1$ and infer $g=\tanh u$ for $K=-1$.

---

[i] Had Lightman et al proceeded the other way around, carrying out the calculation explicitly only for $K=-1$ and then extended the solution to cover $K=+1$, the exponential solution of (28) would have been available to them at this point. But if they had proceeded to (31) with just $K=-1$, they would have had to extract the exponential solution from the hyperbolic transformation by the limiting procedure given in Sec. A4.

## References

1 S. Weinberg, *Gravitation and Cosmology* (John Wiley & Sons, Inc., New York, 1972).
2 N. D. Birrell and P. C. W. Davies, *Quantum fields in curved space* (Cambridge University Press, Cambridge, 1982).
3 R. M. Wald, *General Relativity* (University of Chicago Press, Chicago, 1984).
4 R. C. Tolman, *Relativity Thermodynamics and Cosmology* (Oxford University Press, Oxford, UK, 1934).
5 S. W. Hawking and G. F. R. Ellis, *The Large Scale Structure of Space-Time* (Cambridge University Press, Cambridge, 1973).
6 C. W. Misner, K. S. Thorne, and J. A. Wheeler, *Gravitation* (W. H. Freeman and Co., New York, 1973).
7 B. F. Schutz, *A First Course in General Relativity* (Cambridge University Press, Cambridge, 1985).
8 P. J. E. Peebles, *Principles of Physical Cosmology* (Princeton University Press, Princeton, NJ, 1993).
9 E. R. Harrison, *Cosmology* (Cambridge University Press, Cambridge, UK, 2000).
10 M. Roos, *Introduction to Cosmology* (Wiley, Chichester, UK, 2003).
11 S. Dodelson, *Modern Cosmology* (Academic Press, London, 2003).
12 S. Carroll, *Spacetime and Geometry: An introduction to General Relativity* (Addison-Wesley, San Francisco, 2004).
13 M. P. Hobson, G. Efstathiou, and A. Lasenby, *General Relativity: An Introduction for Physicists* (Cambridge University Press, Cambridge, 2006).
14 L. Bergström and A. Goobar, *Cosmology and Particle Astrophysics* (Springer, 2004).
15 L. D. Landau and E. M. Lifshitz, *The Classical Theory of Fields* (Pergamon Press, Oxford, UK, 1980), Vol. 2.
16 J. A. Peacock, *Cosmological Physics* (Cambridge University Press, Cambridge, UK, 1999).
17 A. P. Lightman, W. H. Press, R. H. Price, and S. A. Teukolsky, *Problem Book in Relativity and Gravitation* (Princeton University Press, Princeton, 1975).
18 H. Stephani *et al.*, in *Exact Solutions of Einstein's Field Equations*, (Cambridge University Press, Cambridge, 2003), Chap. 14, p. 210.
19 G. E. Tauber, J. Math. Phys. **8,** 118 (1966).
20 H. Stephani, *Relativity: An Introduction to Special and General Relativity* (Cambridge University Press, Cambridge, 2004).
21 L. Infeld and A. Schild, Phys. Rev. **68,** 250 (1945).
22 A. G. Walker, M. Not. R. Ast. Soc. **95,** 263 (1935).
23 L. Infeld and A. Schild, Phys. Rev. **70,** 410 (1946).
24 G. Endean, Astrophys. J. **434,** 397 (1994).
25 G. Endean, M. Not. R. Ast. Soc. **277,** 627 (1995).
26 G. Endean, Astrophys. J. **479,** 40 (1997).
27 G. Endean, J. Math. Phys. **39,** 1551 (1998).
28 L. Querella, Astrophys. J. **508,** 129 (1998).
29 A. Herrero and J. A. Morales, J. Math. Phys. **41,** 4765 (2000).
30 C. F. Sopuerta, J. Math. Phys. **39,** 1024 (1998).
31 A. J. Keane and R. K. Barrett, Class. Quantum Grav. **17,** 201 (2000).
32 M. IIhoshi, S. V. Ketov, and A. Morishita, hep-th/0702139v3 (2007) & Prog. Theor. Phys., to be published.
33 H. P. Robertson, Astrophys. J. **82,** 284 (1935).
34 A. G. Walker, Proceedings of the London Mathematical Society (2) **42,** 90 (1936).
35 H. Bondi and T. Gold, M. Not. R. Ast. Soc. **108,** 252 (1948).
36 A. Lasenby and C. Doran, Phys. Rev. D **71,** 063502 (2005).

37 M. Ibison, Class. Quantum Grav. **23,** 577 (2006).
38 A. Lasenby and C. Doran, in *Conformal models of de Sitter space, initial conditions for inflation and the CMB*, edited by C. J. A. P. Martins *et al.* (AIP, 2004), p. 53.
39 A. Lasenby, Phil. Trans. Roy. Soc., to be published.
40 H. Bondi, *Cosmology* (University Press, Cambridge, 1961).
41 E. A. Milne, *Kinematic Relativity* (Oxford University Press, Oxford, UK, 1948).
42 S. Gull, A. Lasenby, and C. Doran, Found. Phys. **23,** 1329 (1993).
43 J. Aczél, *Lectures on Functional Equations and Their Applications* (Dover, Mineola, New York, 2006).

**Figure I caption**

The set of Robertson-Walker spacetimes is the intersection between the set of two-parameter conformal spacetimes and the set of orthogonal time isotropic spacetimes.

The $K$ = -1 hyperbolic space Robertson-Walker spacetimes are the same as the set of conformally-represented spacetimes whose conformal factor is a function of $x^2$ only. These spacetimes contain all 3 maximally symmetric spacetimes.

The de Sitter spacetime is the intersection of all three Robertson-Walker spacetimes.

The Minkowski spacetime is the intersection of the $K$ = -1 and the $K$ = 0 Robertson-Walker spacetimes with $K$ = +1 excluded.

The anti-de Sitter spacetime is a member only of the $K$ = -1 hyperbolic space Robertson-Walker spacetimes.

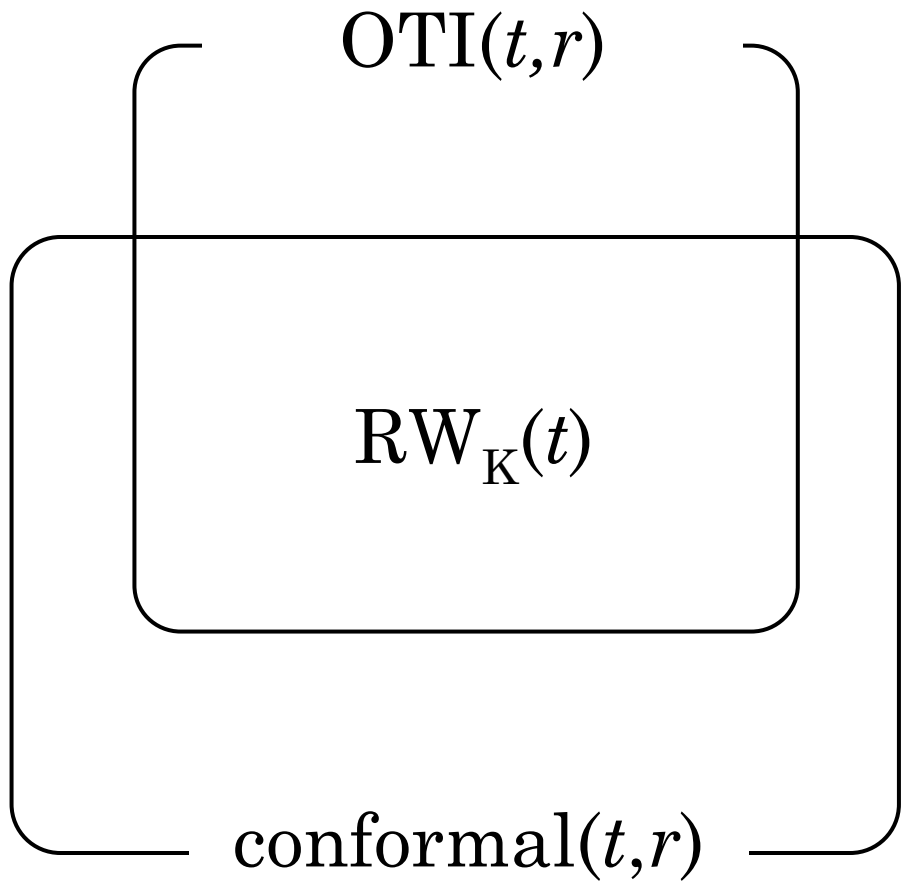
OTI($t$,$r$)
RW$_K$($t$)
conformal($t$,$r$)

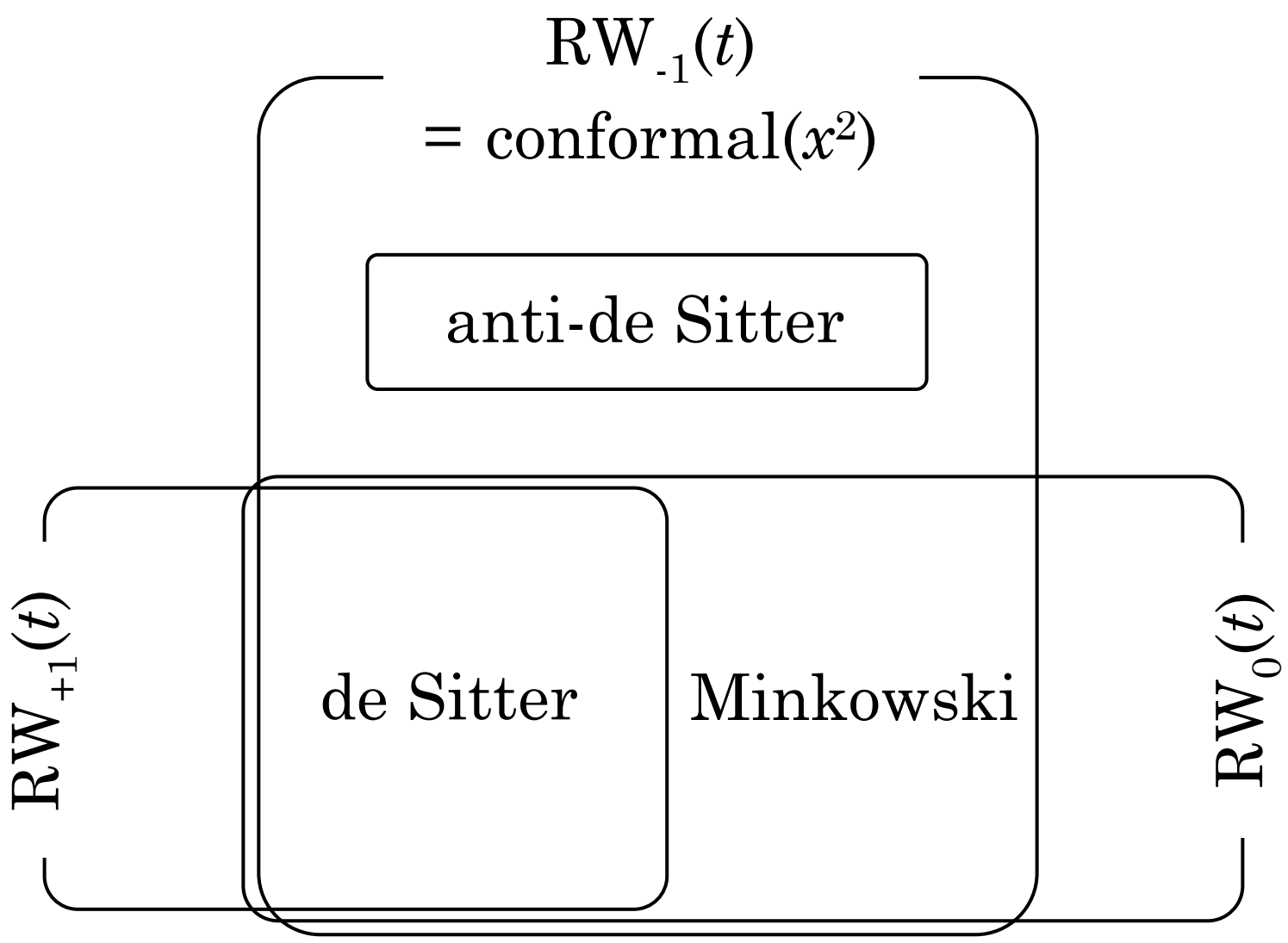